\documentclass[twocolumn]{aastex62}
\usepackage{natbib}
\graphicspath{{./}{figures/}}
\usepackage{color,subfigure,lineno} 

\newcommand{\SiII}{\makebox{[Si{\sc\,II}]\,}}

\newcommand{\hbeta}{H{$\beta$}}
\newcommand{\halpha}{H{$\alpha$}}
\newcommand{\lya}{Ly\,$\alpha$}
\newcommand{\CIII}{C\,{\sc iii]}}
\newcommand{\AlIII}{Al\,{\sc iii}}
\newcommand{\SiIII}{Si\,{\sc iii]}}
\newcommand{\CaII}{Ca\,{\sc ii}}
\newcommand{\OII}{[O\,{\sc ii}]}
\newcommand{\NeV}{[Ne\,{\sc v}]}
\def\CIV{C\,{\sc iv}}
\def\MgII{Mg\,{\sc ii}}
\def\FeII{Fe\,{\sc ii}}
\def\OIII{[O\,{\sc iii}]\,5007}
\def\CaII{Ca\,{\sc ii}}
\def\HeII{He\,{\sc ii}\,1640}
\def\SiIV{Si\,{\sc iv}}

\shorttitle{Spectral Measurements of DR16Q Quasars}
\shortauthors{Wu \& Shen}

\begin{document}

\title{A Catalog of Quasar Properties from Sloan Digital Sky Survey Data Release 16}


\author[0000-0003-4202-1232]{Qiaoya Wu}
\email{qiaoyaw2@illinois.edu}
\affiliation{Department of Astronomy, University of Illinois at Urbana-Champaign, Urbana, IL 61801, USA}
\affiliation{Center for AstroPhysical Surveys, National Center for Supercomputing Applications, University of Illinois at Urbana-Champaign, Urbana, IL 61801, USA}

\author[0000-0003-1659-7035]{Yue Shen}
\email{shenyue@illinois.edu}
\affiliation{Department of Astronomy, University of Illinois at Urbana-Champaign, Urbana, IL 61801, USA}
\affiliation{National Center for Supercomputing Applications, University of Illinois at Urbana-Champaign, Urbana, IL 61801, USA}

\begin{abstract}
We present a catalog of continuum and emission line properties for 750,414 broad-line quasars included in the Sloan Digital Sky Survey Data Release 16 quasar catalog (DR16Q), measured from optical spectroscopy. These quasars cover broad ranges in redshift ($0.1\lesssim z\lesssim 6$) and luminosity ($44\lesssim \log (L_{\rm bol}/{\rm erg\,s^{-1}})\lesssim 48$), and probe lower luminosities than an earlier compilation of SDSS DR7 quasars. Derived physical quantities such as single-epoch virial black hole masses and bolometric luminosities are also included in this catalog. We present improved systemic redshifts and realistic redshift uncertainties for DR16Q quasars using the measured line peaks and correcting for velocity shifts of various lines with respect to the systemic velocity. About 1\%, 1.4\%, and 11\% of the original DR16Q redshifts deviate from the systemic redshifts by $|\Delta V|>1500\,{\rm km\,s^{-1}}$, $|\Delta V|\in [1000,1500]\,{\rm km\,s^{-1}}$, and $|\Delta V|\in [500,1000]\,{\rm km\,s^{-1}}$, respectively; about $1900$ DR16Q redshifts were catastrophically wrong ($|\Delta V|>10,000\,{\rm km\,s^{-1}}$). We demonstrate the utility of this data product in quantifying the spectral diversity and correlations among physical properties of quasars with large statistical samples. 
\end{abstract}


\section{Introduction}\label{sec:intro}

Wide-field spectroscopic surveys such as the Sloan Digital Sky Survey have greatly increased the inventory of quasars over broad ranges in redshift and luminosity \citep[e.g.,][]{Schneider_etal_2010,Shen_etal_2011,Dawson_etal_2013,Paris_etal_2017,Lyke_etal_2020}. Optical spectra from these surveys provide rich information on the physical properties of these quasars, enabling detailed statistical studies of their abundance, accretion parameters, and correlations among physical properties. 

Using optical spectroscopy from the SDSS I-II legacy survey, \citet{Shen_etal_2011} measured spectral properties for 105,783 quasars included in SDSS DR7. Such spectral measurements have been extended to later SDSS quasar catalogs \citep[e.g.,][]{Rakshit_etal_2020} for larger samples of quasars observed in SDSS-III \citep{Eisenstein_etal_2011} and SDSS-IV \citep{Blanton_etal_2017}. These later generations of SDSS surveys target quasars to fainter magnitudes than the SDSS-I/II survey, with upgraded SDSS optical spectrographs \citep{Smee_etal_2013}, thus expanding the dynamical range in SDSS quasar samples. 

In this work, we measure spectral properties for the $\sim 750$k quasars included in the latest SDSS DR16 quasar catalog \citep{Lyke_etal_2020}. As a major update to the SDSS DR7 quasar catalog presented in \citet{Shen_etal_2011}, we present measurements for a more comprehensive list of emission lines and refined systemic redshifts based on these spectral measurements. We also compile derived quantities including black hole mass and Eddington ratio estimates to facilitate statistical studies of SDSS quasars. This paper is organized as follows. In \S\ref{sec:data} we describe the input data of the catalog. We detail our spectral measurements in \S\ref{sec:analysis} and describe the compiled properties in \S\ref{sec:cat}. We present example applications of this catalog in \S5 and conclude in \S\ref{sec:con}. Throughout this paper, we adopt a flat $\Lambda$CDM cosmology with $\Omega_{\Lambda}=0.7$, $\Omega_{M}=0.3$ and $H_0=70\,{\rm km\,s^{-1}Mpc^{-1}}$. 


\section{Data}\label{sec:data}


\begin{figure}
\centering
    \includegraphics[width=\linewidth]{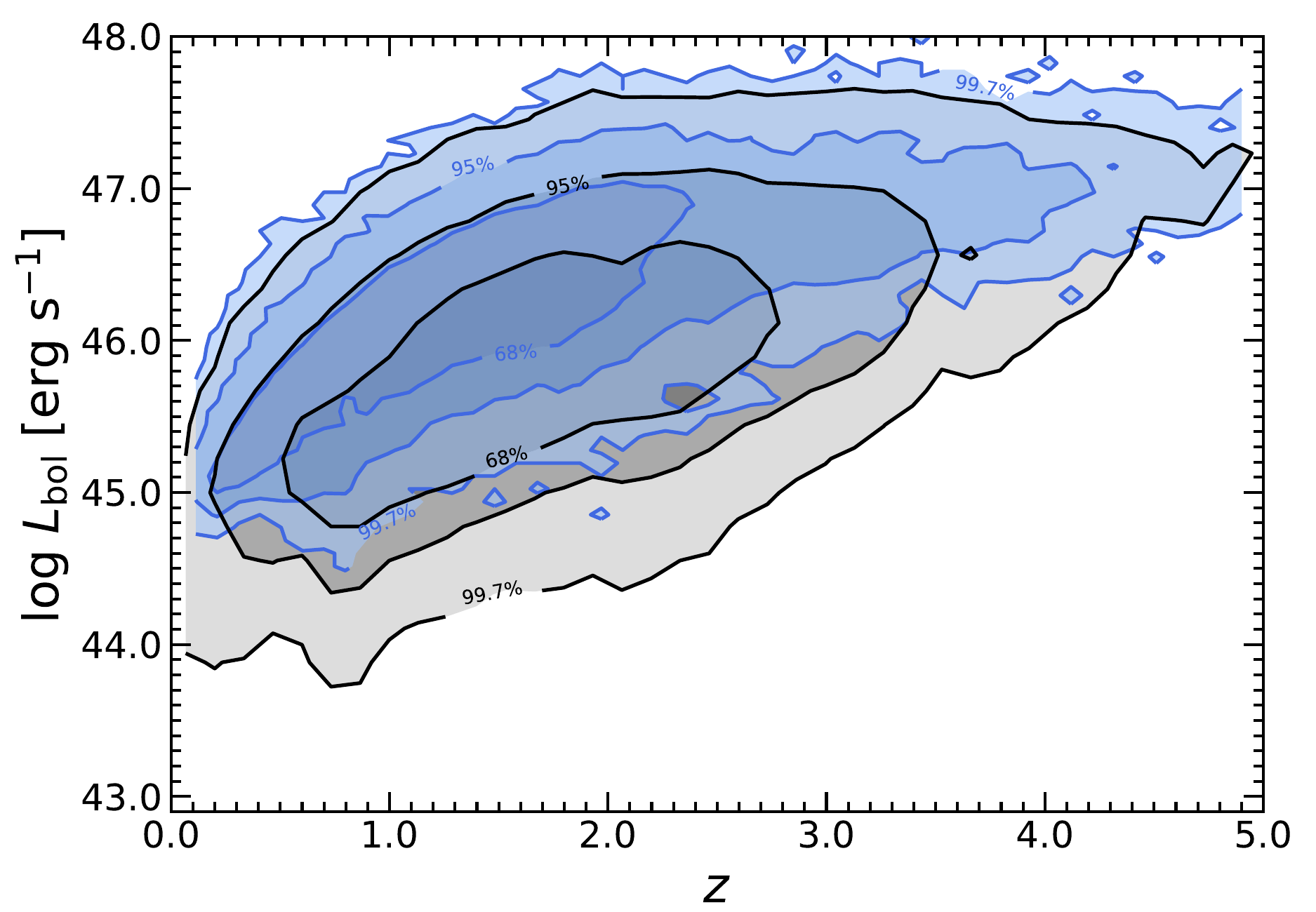}
    \caption{Distribution of quasars in the redshift and luminosity space. The DR16Q sample is shown as black contours. The blue contours are for the SDSS DR7 quasars compiled in \citet{Shen_etal_2011}. Enclosed percentiles are marked for each contour level. } 
    \label{fig:z_Lbol}
\end{figure}

\begin{figure}
\centering
    \includegraphics[width=\linewidth]{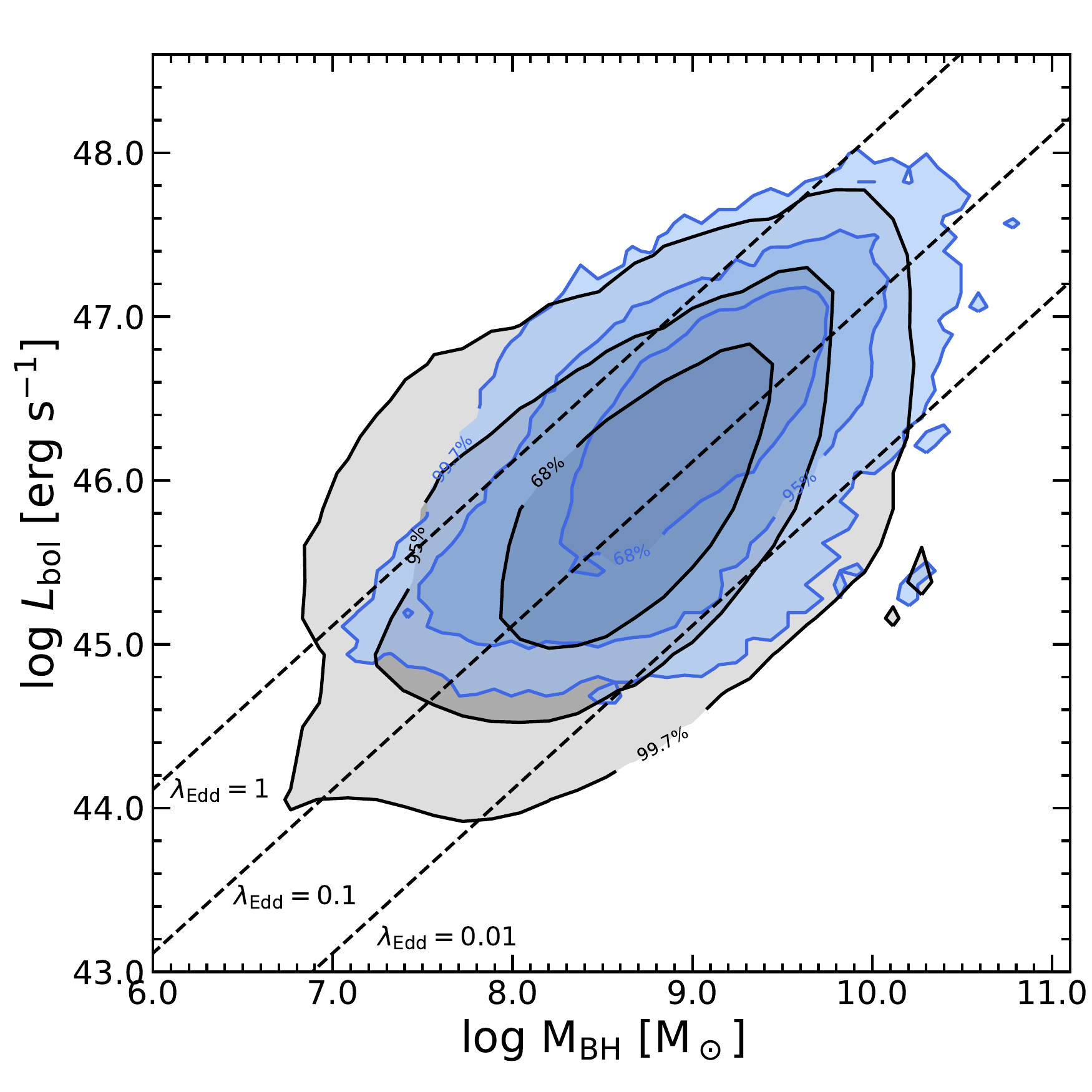}
    \caption{Distribution of quasars in the BH mass and bolometric luminosity space. The DR16Q sample is shown as black contours. The blue contours are for the SDSS-DR7 quasars compiled in \citet{Shen_etal_2011}. Enclosed percentiles are marked for each contour level. We caution that quasars with apparent Eddington ratios $\lambda_{\rm Edd}>1$ are likely caused by uncertainties in the BH mass estimates.}
    \label{fig:Lbol_Mbh}
\end{figure}

\begin{figure*}
\centering
    \includegraphics[width=\linewidth]{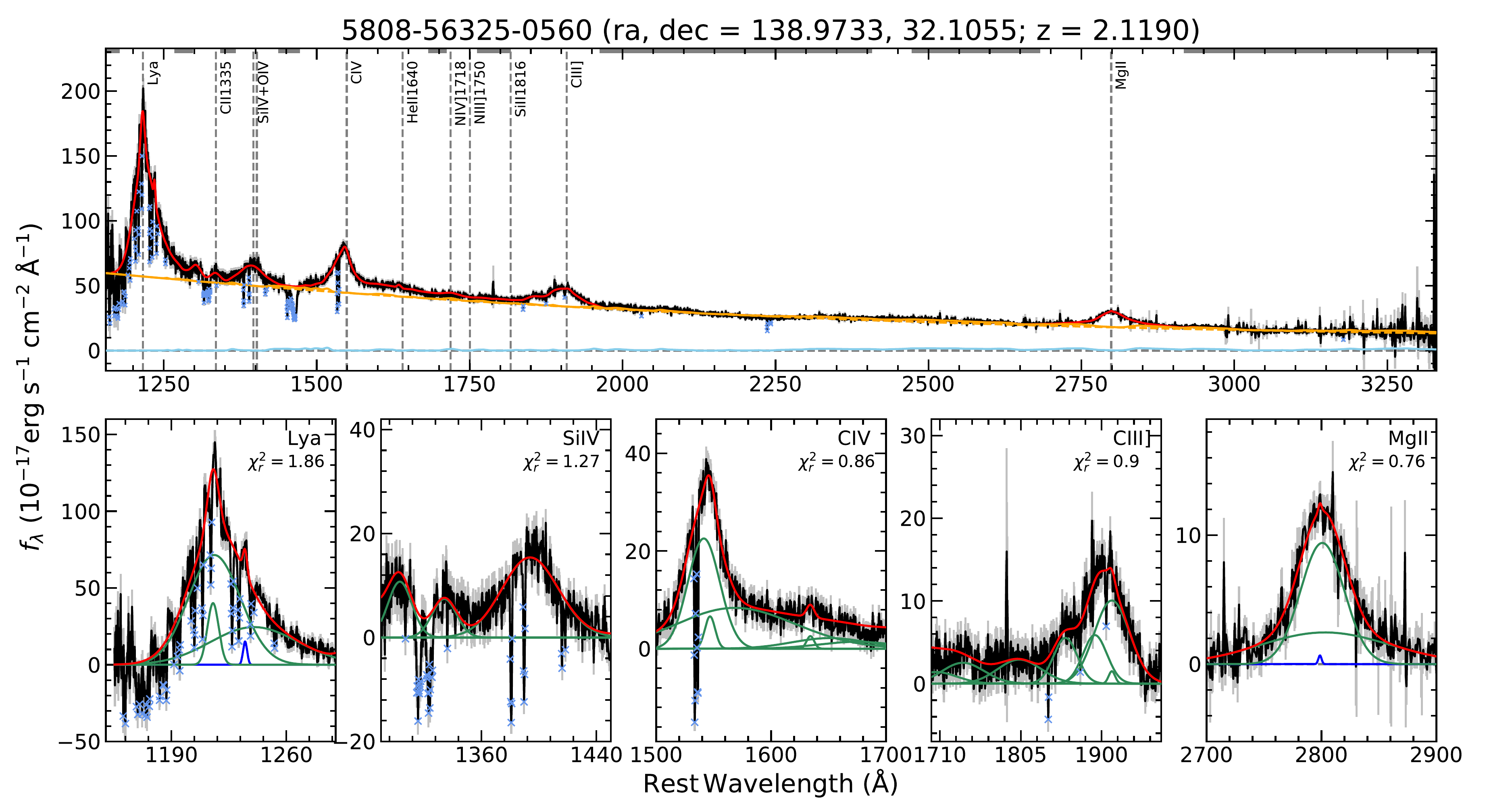}
    \caption{An example of quality assessment (QA) plot for our global spectral fitting approach. The top panel shows the continuum (orange) and \FeII\ (sky-blue) model components; the red line is the global model combining all continuum and emission line components. The light blue crosses are pixels within the continuum and emission-line-fitting windows that are masked as absorption and excluded from the model fit. The gray line segments near the top of the panel indicate the windows used for the continuum + \FeII\, fit. The bottom panels present the emission-line fits for several line complexes, where the green (blue) lines represent broad (narrow) Gaussian components within that line complexes.}
    \label{fig:QA}
\end{figure*}


We start from the DR16Q catalog, which includes 750,414 quasars observed through SDSS-I/II/III/IV. This quasar catalog represents the final quasar sample from SDSS-IV (i.e., SDSS-IV DR17 does not contain new quasars). More than half of these quasars are nearly discovered in SDSS-IV. For completeness, DR16Q also includes quasars observed in previous generations of SDSS, specifically quasars from legacy SDSS-I/II \citep[DR7Q,][]{Schneider_etal_2010} and SDSS-III \citep[DR12Q,][]{Paris_etal_2017}. If a quasar is re-observed in SDSS-IV, then the SDSS-IV observation is taken as the primary spectroscopy for that object. Cross-matching sources in DR16Q with the DR7 quasar catalog \citep{Schneider_etal_2010}, there are 85 bona fide quasars included in DR7Q that are missing from DR16Q. We do not attempt to recover these 85 quasars since their measurements are included in \citet{Shen_etal_2011}. For initial redshift used for our spectral fitting, we adopt the best redshift in DR16Q \citep[Column `Z' in][]{Lyke_etal_2020} and denote it as $z_{\rm DR16Q}$. 

SDSS optical spectroscopy has been obtained with the SDSS-I/II spectrographs for DR7 data and the upgraded BOSS spectrographs \citep{Smee_etal_2013} for SDSS-III and SDSS-IV data. The characteristics of these spectroscopic data are similar but differ in details: DR7 spectra have a wavelength coverage of $\sim 3800-9200\,$\AA\ while BOSS spectra have a wavelength coverage of $\sim 3650-10400\,$\AA. The spectral resolution is roughly the same, $R\sim 2000$ and both SDSS-I/II and SDSS-III/IV spectra are stored in vacuum wavelength with a pixel dispersion of $69\,{\rm km\,s^{-1}}$ (i.e., the wavelength array is logarithmically binned with a dispersion of $d\log_{10}\lambda=10^{-4}$). The BOSS spectrographs accommodate a total of 1000 fibers ($2\arcsec$ in diameter) while the legacy SDSS spectrographs accommodated 640 fibers ($3\arcsec$ in diameter). In addition, BOSS spectrographs have improved throughput than the legacy SDSS spectrographs, resulting in higher S/N at fixed magnitude and exposure time than legacy SDSS I-II spectra.   

In Fig.~\ref{fig:z_Lbol} and Fig.~\ref{fig:Lbol_Mbh} we show the distributions of DR16Q quasars in the redshift-luminosity plane and the black hole mass-luminosity plane, using spectral measurements described in \S\ref{sec:analysis}. Compared with quasars in the DR7Q catalog, DR16Q quasars probe fainter luminsoities, given improved sensitivity of the BOSS spectrographs that allowed targeting to fainter flux limits. 

\section{Spectral Measurements}\label{sec:analysis}

We follow the practice in earlier work \citep[e.g.,][]{Shen_etal_2016b,Shen_etal_2019b} to fit the SDSS spectra with a global continuum+emission lines model, using the public code PyQSOFIT \citep[e.g.,][]{Guo_etal_2018,Shen_etal_2019b} developed by our group. Minor custom adjustments of the fitting constraints are made to produce more robust spectral measurements of DR16Q quasars. The input fitting parameter file is provided in the supplemental materials\footnote{\url{https://github.com/QiaoyaWu/sdss4_dr16q_tutorial}}.

The detailed methodology of our spectral fitting is described in earlier work \citep{Shen_etal_2016b,Shen_etal_2019b,Guo_etal_2018}. In brief, we first correct the observed spectrum for Milky Way reddening using the dust map from \citet{SFD} (updated to the values in \citet{Schlafly_Finkbeiner_2011}) and the \citet{CCM} extinction curve with $R_V=3.1$. The observed spectrum is then shifted to the quasar rest-frame using the default redshift in DR16Q to be fitted by the model.

The continuum model fits the dereddened, de-redshifted spectrum in several relatively line-free regions with the combination of a power law and a third-order polynomial. These continuum-fitting windows are listed in Table \ref{tab:contifit_window}. The additive (positive-definite) polynomial component is introduced to fit objects with peculiar continuum shapes, e.g., due to peculiar intrinsic dust reddening \citep[e.g.,][]{Shen_etal_2019b}. In addition to the power-law and polynomial components, we include the optical and UV \FeII\ emission using empirical templates in the same continuum-fitting windows \citep[][]{Boroson_Green_1992,Vestergaard_Wilkes_2001, Tsuzuki_etal_2006,Salviander_etal_2007}. The continuum and the \FeII\ emission form a pseudo-continuum, which we subtract from the original spectrum to form a line-only spectrum. We then fit the line-only spectrum with a set of Gaussians in logarithmic wavelength space over several line complexes. The details of the line complexes in the fit are summarized in Table \ref{tab:linefit_para}. 



\begin{flushleft}
\begin{table}
\caption{Rest-frame Continuum Fitting Windows}\label{tab:contifit_window}
\scalebox{1.0}{
\begin{tabular}{cc}
\hline\hline
Continuum windows & Fitting range [\AA] \\
(1) & (2) \\
\hline
1 & 1150-1170 \\
2 & 1275-1290 \\ 
3 & 1350-1360 \\
4 & 1445-1465 \\
5 & 1690-1705 \\
6 & 1770-1810 \\
7 & 1970-2400 \\ 
8 & 2480-2675 \\ 
9 & 2925-3400 \\ 
10 & 3775-3832 \\ 
11 & 4000-4050 \\ 
12 & 4200-4230 \\
13 & 4435-4640 \\ 
14 & 5100-5535 \\ 
15 & 6005-6035 \\ 
16 & 6110-6250 \\ 
17 & 6800-7000 \\ 
18 & 7160-7180 \\
19 & 7500-7800 \\ 
20 & 8050-8150 \\
\hline
\hline\\
\end{tabular}
}
\end{table}
\end{flushleft}


\begin{flushleft}
\begin{table}
\caption{Line Fitting Parameters}\label{tab:linefit_para}
\scalebox{0.8}{
\begin{tabular}{ccccc}
\hline\hline
Complex & Fitting range [\AA] & $N_{\rm pix,max}$ & Line & $n_{\rm gauss}$ \\
(1) & (2) & (3) & (4) & (5) \\
\hline
\halpha & 6400-6800 & 264 &  broad \halpha & 3 \\
  & & & narrow \halpha & 1 \\
  & & & [NII]6549   & 1 \\
  & & & [NII]6585   & 1 \\
  & & & [SII]6718   & 1 \\
  & & & [SII]6732   & 1 \\
\hbeta  & 4640-5100 & 411 & broad \hbeta & 3 \\
  & & & narrow \hbeta & 1 \\
  & & & [OIII]4959 core &  1 \\
  & & & [OIII]5007 core & 1 \\
  & & & [OIII]4959 wing & 1 \\
  & & & [OIII]5007 wing & 1 \\
  & & & broad He{\sc ii} 4687 & 1 \\
  & & & narrow He{\sc ii}  4687 & 1 \\
\MgII & 2700-2900 & 311 & broad \MgII\ & 2 \\
  & & & narrow \MgII\ & 1 \\
\CIII\ & 1700-1970 & 641 & \CIII & 2 \\
 & & & \SiIII\,1892 & 1 \\
 & & & \AlIII\,1857 & 1 \\
 & & & \SiII\,1816 & 1 \\
 & & & N{\sc iii}  1750 & 1 \\
 & & & N{\sc iv} 1718 & 1 \\
\CIV & 1500-1700 & 544 & \CIV & 3 \\
 & & & broad He{\sc ii}  1640 & 1 \\
 & & & narrow He{\sc ii}  1640 & 1 \\
 & & & broad O{\sc iii}  1663 & 1 \\
 & & & narrow O{\sc iii}  1663 & 1 \\
\SiIV & 1290-1450 & 508 & broad SiIV/OIV] & 2 \\
 & & & C{\sc ii}  1335 & 1 \\
 & & & O{\sc i}  1304 & 1 \\
\lya & 1150-1290 & 499 & \lya\ & 3 \\
 & & & N{\sc v}  1240 & 1 \\
\hline
\hline\\
\end{tabular}
}
\end{table}
\end{flushleft}


As discussed in \citet{Shen_etal_2019b}, we do not include additional model components, such as the Balmer continuum and host galaxy emission, in our spectral fits. This is because the limited S/N and spectral coverage of our spectra often do not allow more sophisticated model fits and will return unreliable results for the majority of quasars with low-to-moderate spectral S/N. Nevertheless, the PyQSOFIT code can turn on these optional model components in the fit.  

For several narrow emission (stellar absorption) lines of interest, our global model often does not provide an accurate estimate of the local continuum level around them, resulting in biased line measurements. To remedy this, we re-fit these emission/absorption features with a local continuum+line model. The details of the local fits around these lines are summarized in Table \ref{tab:linefit_para_local}. 

Finally, high-redshift quasar spectra are often affected by intervening or intrinsic absorption lines. Following earlier work \citep[e.g.,][]{Shen_etal_2019b}, we remedy the effect of absorption lines with an iterative approach to reject pixels fall $3\sigma$ below the original model fit and refit. This automated absorption-line mitigation scheme does a good job reducing the impact from narrow absorption features in the fit, but improvement in objects affected by broad absorption troughs is only mild. 


\begin{table}
\caption{Local Line Fitting Parameters}\label{tab:linefit_para_local}
\scalebox{0.8}{
\begin{tabular}{ccccc}
\hline\hline
Complex & Fitting range [\AA] & $N_{\rm pix, max}$ & Line &  $n_{\rm gauss}$ \\
(1) & (2) & (3) & (4) & (5) \\
\hline
\CaII & 3900-3960 & 67  & \CaII\,3934 & 2 \\
\OII  & 3650-3800 & 175 & \OII\,3728 & 1 \\
\NeV  & 3380-3480 & 127 & \NeV\,3426 core & 1 \\
      &           &     & \NeV\,3426 wing & 1 \\
\hline
\hline\\
\end{tabular}
}
\end{table}

We measure the continuum and emission-line properties using the best-fit model parameters. To estimate the uncertainties of these spectral measurements, we use a Monte Carlo approach \citep[e.g.,][]{Shen_etal_2011,Shen_etal_2019b}: the original spectrum is perturbed by adding a Gaussian random deviate at each pixel using the reported spectral flux errors to generate a mock spectrum; we fit the mock spectra from 25 random trials and take the semi-amplitude of the 16th and 84th percentile range as the measurement uncertainty in the reported spectral property. We have tested with more random trials (e.g., 100) for a random subset of quasars and did not find significant differences in the uncertainty estimates. We therefore adopt 25 Monte Carlo trials per quasar to speed up the spectral fits.

\section{Catalog Compilation}\label{sec:cat}

We compile our spectral measurements and derived quantities in Table \ref{tab:fits_catalog}. The basic spectral measurements include properties of the continuum (and \FeII\ emission) and prominent broad and narrow emission lines in quasar spectra. Our spectral fits are fully automated with various fitting restrictions to reduce the rate of catastrophic failures of the fitting. Quantities not measured (e.g., not covered by the spectrum) have associated errors $=-1$. Because of the large size of the sample and diverse spectral properties of quasars, there are still cases where the fitting results might be unreliable. We provide all the measured properties, associated uncertainties and quality flags in Table \ref{tab:fits_catalog}.  These reported quantities can be used to select a cleaned sample of measurements. In the case of a specific line, we recommend the following quality cuts: 
\begin{enumerate}
    \item[$\bullet$] line flux / flux error $> 2$\ ,
    \item[$\bullet$] $38< \log (L_{\rm line}/{\rm erg\,s^{-1}})<48$\ ,
    \item[$\bullet$] $N_{\rm pix, line\ complex} > 0.5 \times N_{\rm max}$\ .
\end{enumerate}
Line measurements that do not meet the first cut (i.e., the line is detected at $>2\sigma$) might still be valid, but will be noisy and potentially biased. The last two criteria are particularly useful to remove rare cases where the line is unconstrained due to data gaps in the spectrum, and $N_{\rm max}$ is the maximum number of available pixels in SDSS spectra for a given line complex. 


We also pay specific attention to the \CIV\ fits. Since the \CIV\ FWHM is an essential quantity in the estimation of virial BH masses and we do not impose width constraints in the three Gaussians fit to the entire \CIV\ profile, we reject any Gaussians with fluxes less than $5\%$ of the total \CIV\ flux in measuring the \CIV\ FWHM. This treatment efficiently accounts for cases where one or more Gaussians are adversely fit to noise spikes in the spectrum. The other two broad emission lines used for virial BH mass estimation, \hbeta\ and \MgII, do not require this additional processing as the broad-line profile is usually well fit by the model per our visual inspection.

In addition to these basic spectral measurements, we provide derived quantities, as detailed below. 

\subsection{Bolometric luminosities, black hole masses and Eddington ratios}

We estimate the bolometric luminosity $L_{\rm bol}$ of the quasar using the measured continuum luminosity (\FeII\ emission excluded) at rest-frame wavelengths of 5100\,\AA, 3000\,\AA, and 1350\,\AA, depending on the redshift of the quasar. Bolometric corrections (BC) are derived from the mean spectral energy distribution of quasars in \citet{Richards_etal_2006b}, with ${\rm BC=}9.26, 5.15, 3.81$ at the three wavelengths, respectively. Whenever possible, we use the 3000\,\AA\ continuum luminosity to estimate $L_{\rm bol}$ because it has less host contamination than the 5100\,\AA\ luminosity and suffers less from reddening and variability than the 1350\,\AA\ luminosity. Measurement uncertainties in $L_{\rm bol}$ are propagated from measurement uncertainties in the monochromatic luminosity used. We caution that quasar SEDs can vary significantly among individual objects, and these bolometric luminosity estimates based on a single monochromatic continuum luminosity could have substantially larger systematic uncertainties for individual objects \citep[e.g.,][]{Richards_etal_2006b}. All luminosities are calculated using the \texttt{Z\_FIT} redshift instead of the improved systemic redshifts described in \S\ref{sec:zsys} -- the luminosity differences from this detail are negligible. 

Following previous work \citep[e.g.,][]{Shen_etal_2011,Shen_etal_2019b}, we estimate black hole masses from the continuum and broad emission line measurements. These BH mass estimates are based on the so-called ``single-epoch virial BH mass'' estimators \citep[e.g.,][]{Vestergaard_Peterson_2006}. We adopt the same fiducial BH mass recipes on \hbeta, \MgII\ and \CIV\ as in \citet{Shen_etal_2011,Shen_etal_2019b}. These three BH mass recipes are shown to produce consistent estimates for DR7Q quasars when two lines are available \citep[][and see Fig.~\ref{fig:MBH_comparison}]{Shen_etal_2011}. Alternative mass recipes can be applied using the compiled continuum and broad-line measurements. However, we caution that it is only valid to apply a BH mass recipe that follows the same spectral fitting methodology as adopted here in order to avoid biases in the measured continuum and broad-line properties.  

\begin{figure}
\centering
    \includegraphics[width=\linewidth]{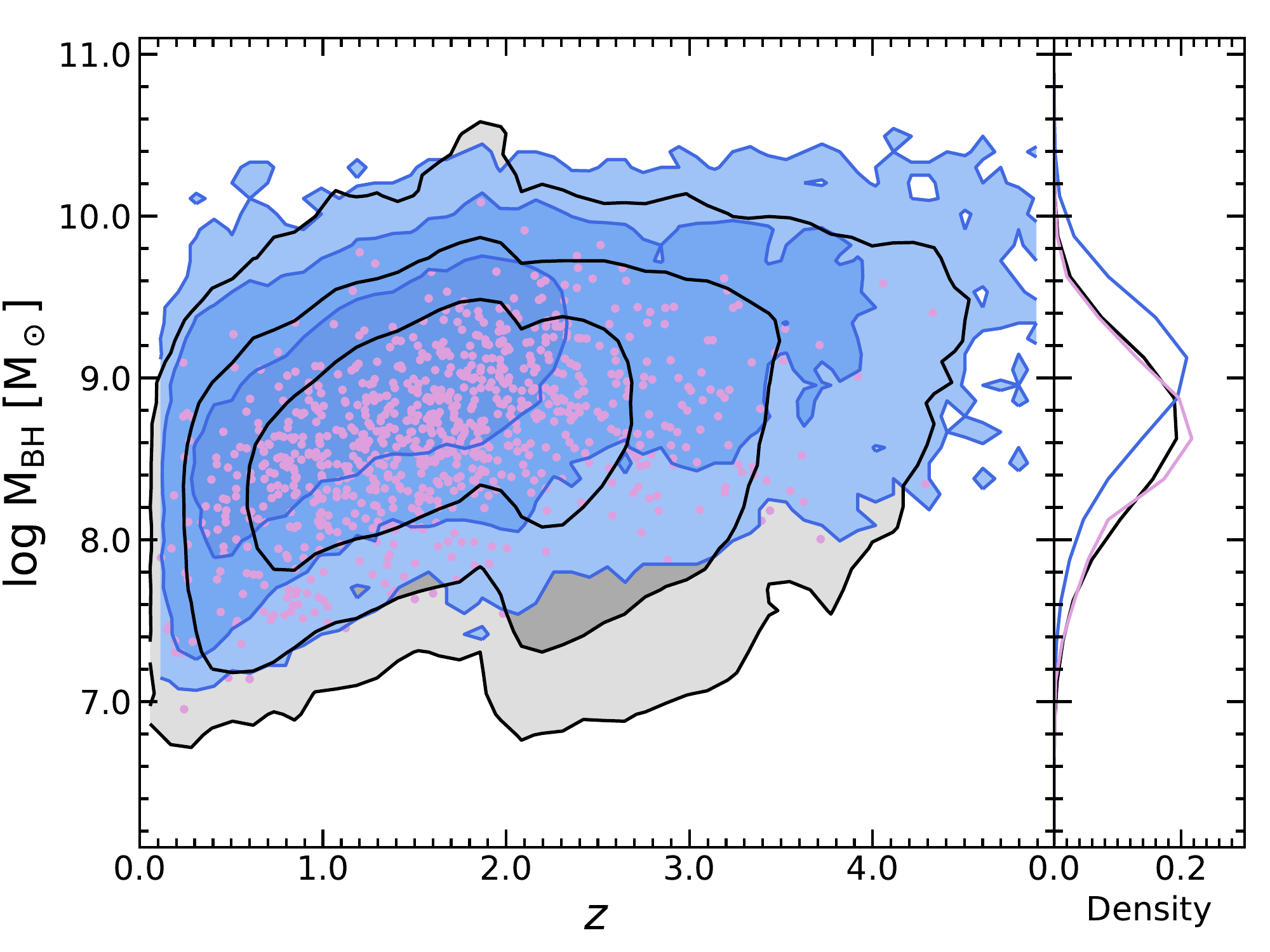}
    \caption{Distribution of quasars in the redshift and BH mass space. The DR16Q sample is shown in black contours; the SDSS-DR7 quasars are shown in blue contours; the SDSS-RM sample \citep{Shen_etal_2019b} is plotted in pink points. The apparent discontinuity around $z\sim 2$ is caused by the switch from \MgII\ BH masses to \CIV\ masses, as discussed in \S\ref{sec:bh_mass}. }
    \label{fig:z_MBH}
\end{figure}

There are significant systematic uncertainties in these single-epoch BH mass estimates \citep[see discussions in, e.g.,][]{Shen_2013}, and we caution on any direct physical interpretations with these virial BH masses. In the catalog, uncertainties in the virial BH masses are measurement uncertainties only, and the systematic uncertainties of these BH masses can be as large as $\sim 0.4$\,dex \citep[e.g.,][]{Shen_2013}.

In the compiled catalog, we provide BH mass estimates based on all three lines (\hbeta, \MgII, and \CIV) whenever available. For the fiducial BH mass, we adopt the \hbeta\ mass for $z<0.7$; for BOSS data, we use \MgII\, and \CIV\, mass for $0.7\leqslant z<2.0$ and $z\geqslant2.0$, respectively; for SDSS legacy data, we use \MgII\, and \CIV\, mass for $0.7\leqslant z<1.9$ and $z\geqslant1.9$, respectively. Fig.~\ref{fig:z_MBH} shows the distribution of DR16Q quasars in the BH mass versus redshift plane. These fiducial BH masses and bolometric luminosities are used to calculate the Eddington ratio $\lambda \equiv L_{\rm bol}/L_{\rm Edd}$, where $L_{\rm Edd}=1.3\times 10^{38} (M_{\rm BH}/M_\odot)\,{\rm erg\,s^{-1}}$ is the Eddington luminosity of the quasar.

\subsection{Improved systemic redshifts}\label{sec:zsys}

Accurate systemic redshift of quasars is crucial to most science applications. As such, extra efforts are often invoked to improve the quasar redshifts produced by the SDSS pipeline \citep[e.g.,][]{Hewett_Wild_2010,Shen_etal_2019b}. The redshift of quasars is mainly constrained by prominent emission line features. However, quasar emission lines are known to have a variety of velocity shifts from the systemic velocity that depend on the line species \citep[e.g.,][]{Gaskell_1982,Tytler_Fan_1992,Richards_etal_2002,Shen_etal_2016b}. \citet{Shen_etal_2016b} studied velocity shifts between different quasar emission lines and \CaII\ absorption (from the host galaxy) using a sample of quasars with high S/N from the coadded BOSS spectra in the SDSS-RM project \citep{Shen_etal_2015a}. Based on these measurements, \citet{Shen_etal_2016b} provided recipes to correct for the average velocity offset from systemic for different lines, as well as realistic uncertainties of the velocity correction. Using this scheme and our line measurements for DR16Q quasars, we derive improved systemic redshifts for all DR16Q quasars. 

The default redshifts in DR16Q are mostly based on the redshifts derived by the SDSS pipeline \citep{Bolton_etal_2012}. \citet{Lyke_etal_2020} also visually inspected a small fraction of quasars and updated their default redshifts in DR16Q with visual redshifts. The SDSS pipeline redshifts were based on PCA template fitting to the observed spectra, where the quasar templates were constructed from a fixed benchmark sample of quasars. As detailed in \citet{Shen_etal_2016b}, the PCA template fitting does not fully account for the diversity in velocity shifts of various quasar emission lines, and only correct for the average velocity shift appropriate for the luminosity ranges probed by the benchmark sample. This is particularly a problem for high-redshift quasars where the strong broad \CIV\ line dominates the redshift estimation. The \CIV\ line is generally blueshifted from the systemic velocity, with its blueshift increasing with quasar luminosity \citep[e.g.,][]{Gaskell_1982,Richards_etal_2002,Shen_etal_2016b}. The default SDSS redshift pipeline does not address well this luminosity-dependent velocity shift of \CIV\ (and a few other high-ionization lines showing similar behaviors), and thus would lead to biased redshifts across the luminosity range of quasars. 

\begin{figure*}
\centering
    \includegraphics[width=\linewidth]{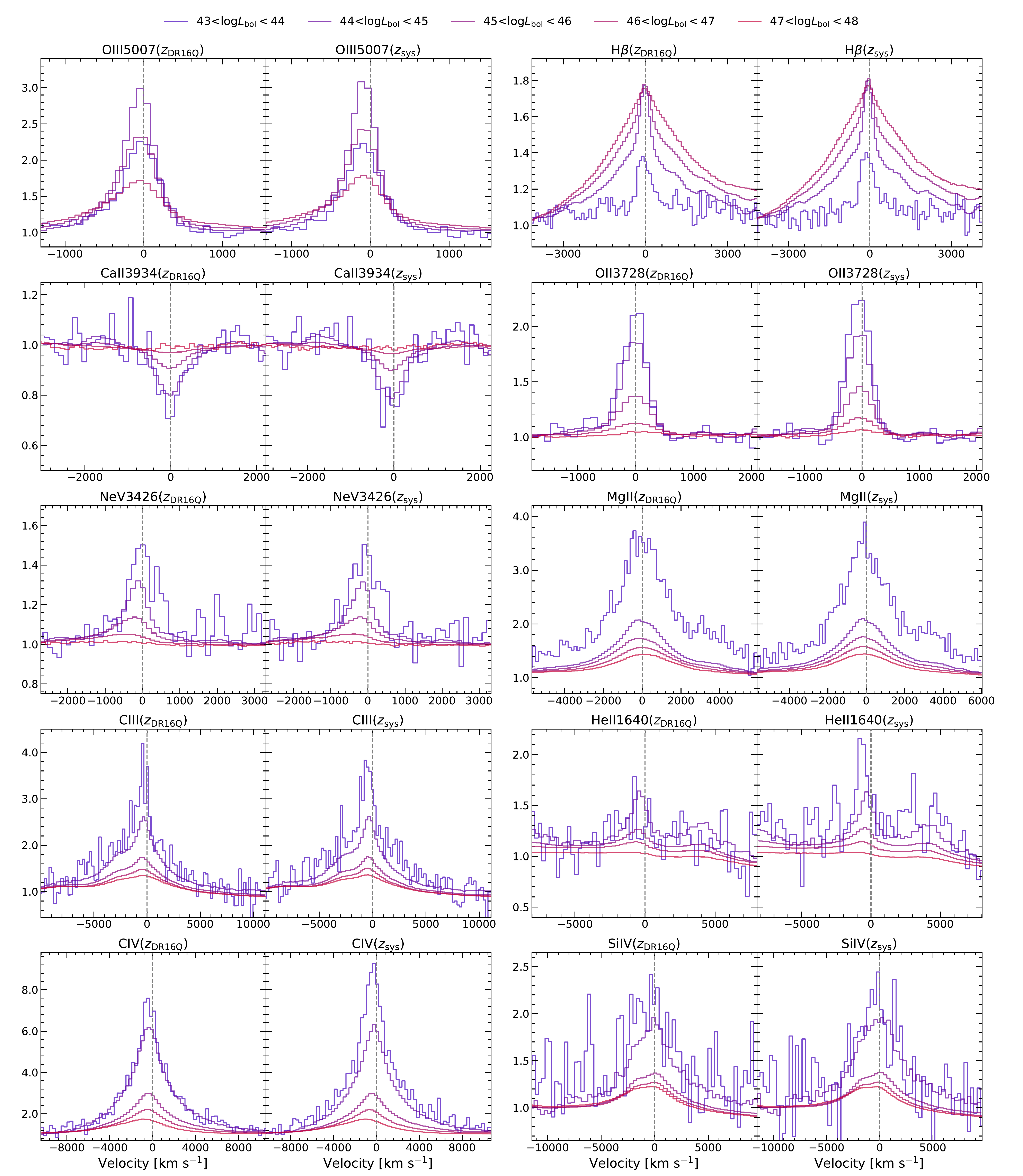}
    \caption{Median composite spectra around several emission/stellar absorption lines of the DR16Q sample, generated using the default DR16Q redshifts (left panels) and the improved systemic redshift (right panels), as a function of quasar luminosity. The improved systemic redshifts make the narrow line features in the composite spectrum sharper, and the peaks of broad lines closer to their expected wavelengths (can be shifted from the systemic velocity).}
    \label{fig:composite_spec}
\end{figure*}

\begin{figure*}
\centering
    \includegraphics[width=\linewidth]{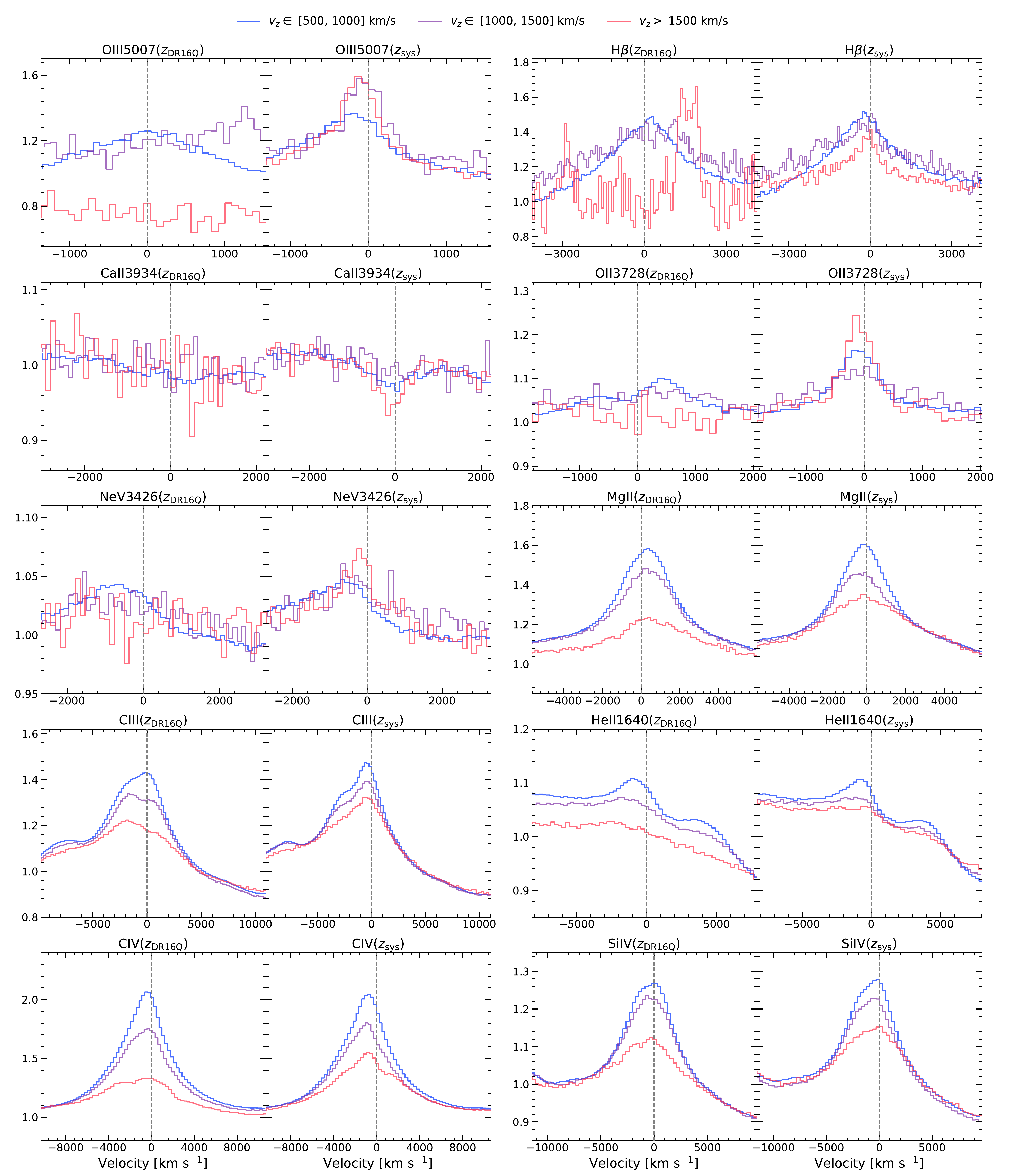}
    \caption{Median composite spectra around several emission/stellar absorption lines of DR16Q quasars with large discrepancies between the default DR16Q redshift and the new systemic redshift, generated using the default DR16Q redshifts (left panels) and the improved systemic redshift (right panels). The new systemic redshifts perform much better than $z_{\rm DR16Q}$ in the statistical sense. }
    \label{fig:z_outlier_composite_spec}
\end{figure*}

\begin{figure}
\centering
    \includegraphics[width=\linewidth]{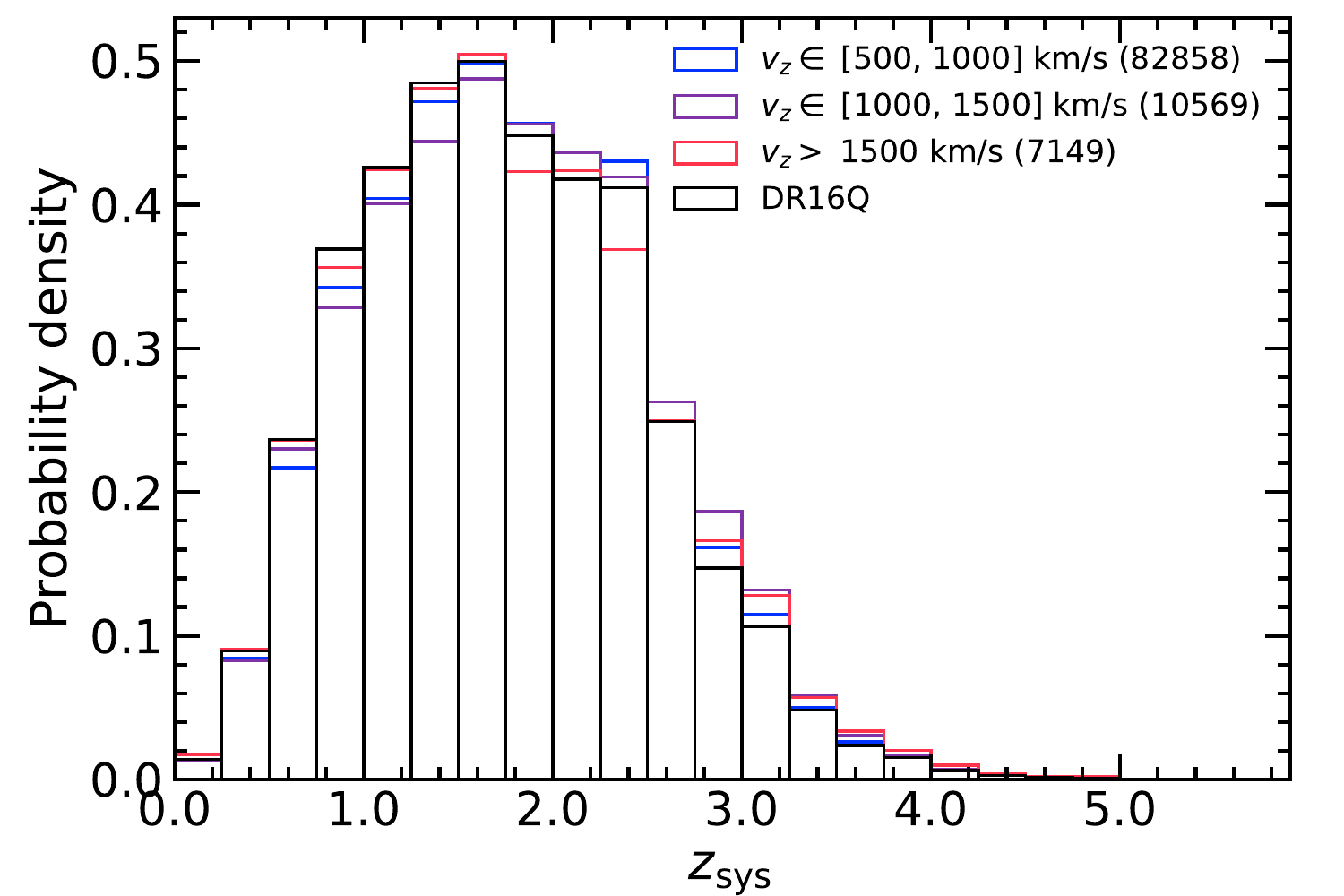}
    \caption{Redshift distributions of DR16Q quasars with large discrepancies between $z_{\rm DR16Q}$ and $z_{\rm sys}$ for the three subsamples shown in Fig.~\ref{fig:z_outlier_composite_spec}. They roughly follow the same redshift distribution of the full DR16Q sample, suggesting no particular redshift failures (of $z_{\rm DR16Q}$) in certain regime. The numbers of quasars in these subsamples are shown in the top-right legend. }
    \label{fig:z_outlier_dist}
\end{figure}

We follow the recipe in \citet{Shen_etal_2016b} to correct for the average intrinsic velocity shifts of different lines, with additional luminosity dependence for three high-ionization lines: \CIV, \HeII\ and \SiIV. However, realizing that the luminosity range probed by DR16Q quasars extends to fainter luminosities than those of the SDSS-RM quasars used in the calibration of velocity shifts in \citet{Shen_etal_2016b}, we force the velocity shifts of \CIV, \HeII\ and \SiIV\ to be zero below $\log (L_{\rm 1700}/{\rm erg\,s^{-1}})=44.5$. This modification is necessary so that we do not artificially introduce redshifted high-ionization lines for the low-luminosity quasars in DR16Q. For secure line redshift measurements, we only use lines detected at $>2\sigma$ and require the spectrum covers at least 50\% of the pixels in the line complex. The individual lines from which we estimate $z_{\rm sys,line}$ are: H$\beta_{\rm br}$, \OIII, \CaII\,3934, \OII\,3728, \NeV\,3426, \MgII, \CIII, \HeII, \CIV, \SiIV. By default the peak from the full line profile is used (broad+narrow component) unless specified otherwise. However, we find many quasars have inaccurate fits for \HeII\ and \NeV\,3426, which would lead to biased redshift estimates. We therefore exclude these two lines in the final systemic redshift estimation below. The $z_{\rm sys,line}$ values for the remaining 8 lines are listed in Table~\ref{tab:fits_catalog}. 


After deriving the systemic redshift $z_{\rm sys,line}$ based on individual lines, we derive a mean $z_{\rm sys, mean}$ from these lines weighted by their individual uncertainties. These redshift uncertainties are the quadratic sum of measurement uncertainties of the line peak and systematic uncertainties from the scatter in the line velocity offset from systemic \citep[e.g.,][]{Shen_etal_2016b}. When calculating the mean systemic redshift using individual $z_{\rm line}$, we reject outlier $z_{\rm line}$ values that are $>3$ times the Median Absolute Deviation from the mean. Compared with individual $z_{\rm sys,line}$, $z_{\rm sys, mean}$ produces the best estimate of the true systemic redshift of the quasar, as at times a particular $z_{\rm line}$ is incorrectly measured due to a bad fit. We demonstrate the performance of $z_{\rm sys,mean}$ with composite spectra (see below). For simplicity, in what follows we use $z_{\rm sys}$ to refer to $z_{\rm sys,mean}$. 

In visual inspection of quasars with redshift differences exceeding $1500\,{\rm km\,s^{-1}}$ between $z_{\rm DR16Q}$ and our $z_{\rm sys}$, we find that many of these quasars have catastrophically incorrect DR16Q redshift. We manually assign redshifts to these quasars, and refit them to derive improved systemic redshifts. The {\tt Z\_FIT} column in the catalog refers to the manual redshift rather than the initial guess $z_{\rm DR16Q}$ for these objects. 

Because we use the peak of the line model to measure systemic redshift following earlier work \citep{Shen_etal_2019b}, we pay specific attention to noisy spikes in the spectrum mistaken as the line peak. This is a problem mainly for \CIV\ and \MgII, where we use the peak from the full model profile (all broad+narrow Gaussians) to measure $z_{\rm sys,line}$. We therefore reject any Gaussians with fluxes less than 5\% of the total line flux before we measure the line peak and $z_{\rm sys,line}$ from it. There is a different rare situation with \CIII, where the peak of the \CIII\ complex is dominated by the adjacent \SiIII\ line. Therefore we use the peak of the \CIII\ line (not the entire \CIII\ complex) to measure $z_{\rm line, CIII]}$.  Although we have applied these additional criteria in measuring the line-based redshift, we preserve the original line peak measurements in the catalog (Table \ref{tab:fits_catalog}). 

We are able to derive systemtic redshift estimates for the vast majority of DR16Q quasars. For the remaining small fraction of objects, the spectra are either too noisy to reliably measure line peaks or the spectrum is the superposition of objects at different redshifts (very rare occasions). For those without $z_{\rm sys}$ measurements, we set $z_{\rm sys}$ to be the input redshift of the fit and its error to $-1$. The redshifts for these objects are highly uncertain and for some of them even the quasar classification is questionable. 


\begin{figure}
\centering
    \includegraphics[width=\linewidth]{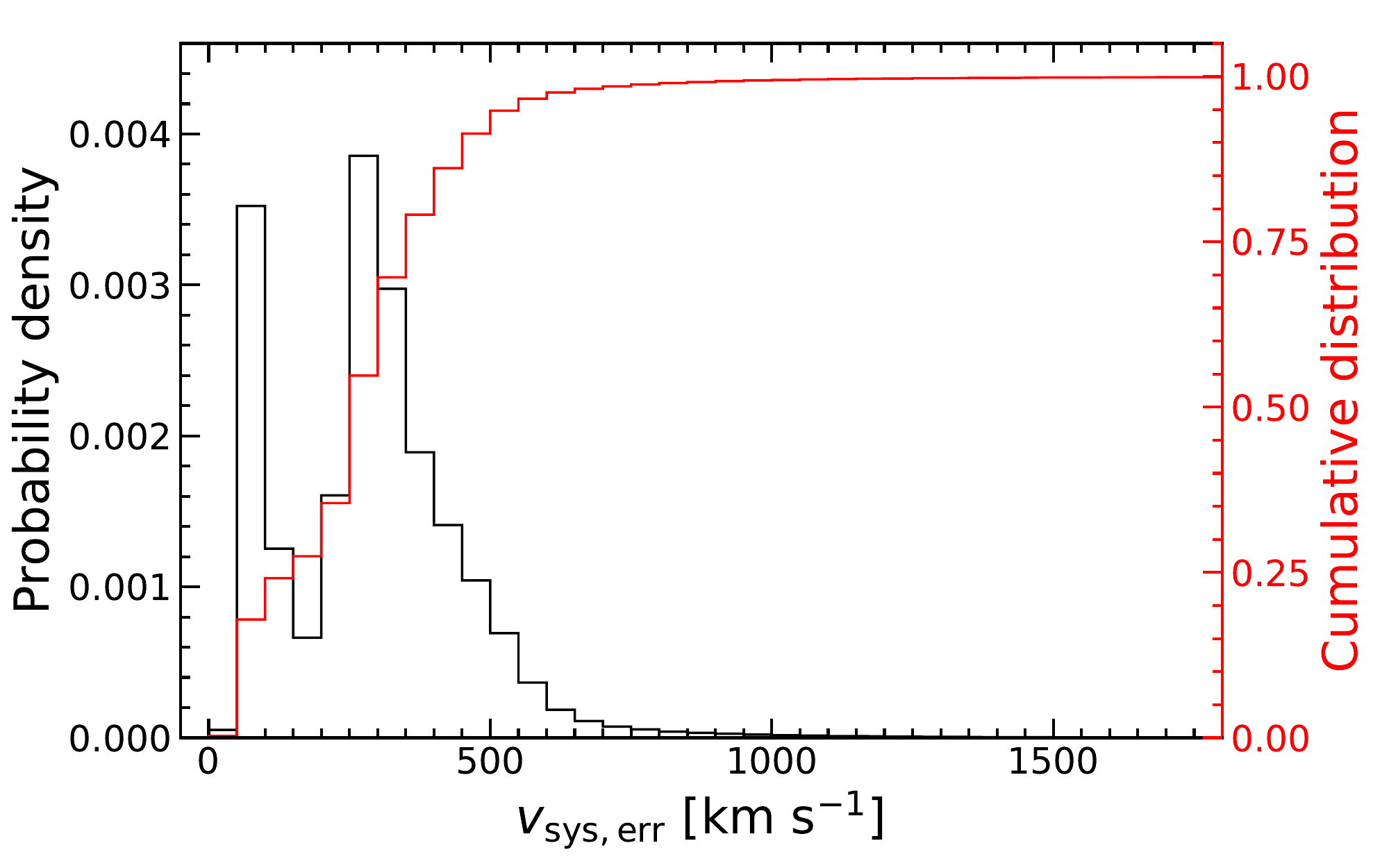}
    \caption{Probability distribution (black) and cumulative distribution (red) of the velocity uncertainty of $z_{\rm sys}$ for the DR16Q sample. }
    \label{fig:vsys_err}
\end{figure}

Fig.~\ref{fig:composite_spec} compares our $z_{\rm sys}$ with $z_{\rm DR16Q}$ using median composite spectra of DR16Q quasars. In this comparison, we divide the sample in different luminosity bins to show any potential luminosity dependence of line shift. For low-ionization broad lines and narrow emission lines, our new redshifts are better than $z_{\rm DR16Q}$ given the slightly sharper profile of the composite line spectrum (in particular for \OII). The peak locations of \OIII\ and \MgII\ are also more consistent with their expected velocity shifts from the systemic velocity \citep{Shen_etal_2016b} using our systemic redshifts. 

Furthermore, for high-ionization broad lines such as \CIV\ with a known luminosity dependence of the line velocity shift, our systemic redshifts reproduce this luminosity trend well. On the contrary, the default DR16Q redshifts failed to produce a clear luminosity trend of \CIV\ blueshift. This is expected because the PCA redshifts are dominated by the mean quasar spectrum (the first PCA eigenvector) from the calibration sample, with a fixed average \CIV\ blueshift.   


We note that for the majority of DR16Q quasars, the difference between $z_{\rm DR16Q}$ and $z_{\rm sys}$ is less than $500\,{\rm km\,s^{-1}}$; thus the extra broadening of the lines due to redshift uncertainties in the composite spectra is more obvious in narrow lines than in broad lines. To further demonstrate the improvement of our redshift estimation, we compare the composite spectra for subsets of DR16Q quasars with large differences between $z_{\rm DR16Q}$ and $z_{\rm sys}$ in Fig.~\ref{fig:z_outlier_composite_spec}. This comparison further demonstrates that our systemic redshifts are substantially better than $z_{\rm DR16Q}$ in the statistical sense. We visually inspect these individual quasars with large redshift discrepancies, and conclude that in the vast majority of cases our systemic redshift is indeed better than $z_{\rm DR16Q}$. On the other hand, there are 141 quasars for which our $z_{\rm sys}$ values are biased by a bad line fit; we fix these 141 quasar redshifts by assigning a manual redshift based on visual inspection and set the $z_{\rm sys}$ uncertainty to be $-2$. 

For $\sim 90\%$ of the DR16Q sample, the redshift difference between $z_{\rm DR16Q}$ and our $z_{\rm sys}$ is small ($|\Delta V|\equiv|c\Delta z/(1+z)|<500\,{\rm km\,s^{-1}}$), indicating that the SDSS redshift pipeline performs well overall. However, 11\%, 1.4\%, and 1\% of the quasars have redshift differences $|\Delta V|\in[500,1000]$, $|\Delta V|\in[1000,1500]$ and $|\Delta V|>1500\,{\rm km\,s^{-1}}$, respectively. We show the redshift distributions of these subsets of quasars in Fig.~\ref{fig:z_outlier_dist}, which show no obvious differences among the three subsets. There are 1943 quasars with $|\Delta V|>10,000\,{\rm km\,s^{-1}}$, in which case we consider $z_{\rm DR16Q}$ to be catastrophically wrong, as resulting from mis-identified lines in the spectrum by the BOSS pipeline. Our visual inspection of the DR16Q sample is not exhaustive and likely even incorrect for rare objects, and therefore there are still a small fraction of quasars with incorrect or inaccurate redshifts. But the fraction of such bad redshifts has been greatly reduced with our approach. In particular, a substantial fraction of quasars with $z_{\rm DR16Q}>5$ turn out to be lower-redshift quasars. 

Finally, in Fig.~\ref{fig:vsys_err} we show the distribution of uncertainties of our systemic redshifts combining measurement and systematic errors. Low-redshift quasars have smaller redshift uncertainties from narrow emission lines, while quasars with broad lines only in their spectra have redshift uncertainties larger than $200\,{\rm km\,s^{-1}}$. The vast majority of quasars have redshift uncertainties less than $500\,{\rm km\,s^{-1}}$.  


\begin{longtable*}{llll}
\caption[notes]{FITS Catalog Format}\label{tab:fits_catalog}\\
\hline \hline \\[0.2ex]
   \multicolumn{1}{l}{\textbf{Column}} &
   \multicolumn{1}{l}{\textbf{Format}} &
   \multicolumn{1}{l}{\textbf{Units}} &
   \multicolumn{1}{l}{\textbf{Description}} \\[0.2ex] \hline
   \\[0.2ex]
\endfirsthead
\multicolumn{4}{l}{{\tablename} \thetable{} -- Continued} \\[0.5ex]
  \hline \hline \\[0.2ex]
  \multicolumn{1}{l}{\textbf{Column}} &
  \multicolumn{1}{l}{\textbf{Format}} &
  \multicolumn{1}{l}{\textbf{Units}} &
  \multicolumn{1}{l}{\textbf{Description}} \\[0.2ex] \hline
  \\[0.2ex]
\endhead
  \hline
  \multicolumn{4}{l}{{Continued on Next Page\ldots}} \\
\endfoot
  \\[0.2ex] \hline \hline
\endlastfoot
 SDSS\_NAME & STRING &    & SDSS DR16 designation (J2000) \\
 PLATE & LONG64 &   & Spectroscopic plate number \\
 MJD & LONG64 &   & Spectroscopic MJD \\
 FIBERID & LONG64 &   & Spectroscopic fiber number \\
 RA & DOUBLE & degree & Right ascension (J2000) \\
 DEC & DOUBLE & degree & Declination (J2000) \\
 OBJID & STRING &    & PLATE-MJD-FIBERID: PyQSOFit output name \\
 IF\_BOSS\_SDSS & STRING &    & Source of the input spectrum: BOSS or SDSS \\
 Z\_DR16Q & DOUBLE &   & Best redshift provided by DR16Q \\
 SOURCE\_Z\_DR16Q & STRING &    &  Source for DR16Q redshift from \citet{Lyke_etal_2020} \\
 Z\_FIT & DOUBLE &   & Input redshift for QSOFit; can differ from Z\_DR16Q \\
 Z\_SYS & DOUBLE &   & Systemic redshift \\
 Z\_SYS\_ERR & DOUBLE &   & Uncertainties of systemic redshift \\
 EBV & DOUBLE &   & Milky Way extinction $E(B-V)$ from \citet{SFD} \\
  & & & and scaled to match the results in \citet{Schlafly_Finkbeiner_2011} \\
 SNR\_MEDIAN\_ALL & DOUBLE &   & Median S/N per pixel of the raw spectrum \\
 CONTI\_FIT & DOUBLE[5] &   & Best-fit parameters for the continuum model (PL+poly) \\
 CONTI\_FIT\_ERR & DOUBLE[5] &   & Uncertainties in the best-fit continuum parameters \\
 CONTI\_FIT\_STAT & DOUBLE[2] &   & Continuum fitting pixel number, reduced $\chi^2$ \\
 FEII\_UV & DOUBLE[3] &   & Best-fit parameters for the UV \FeII\ model \\
 FEII\_UV\_ERR & DOUBLE[3] &   & Uncertainties in the best-fit UV \FeII\ model \\
 FEII\_UV\_EW & DOUBLE & $\rm \AA$ & Rest-frame equivalent width of UV \FeII\ within 2250-2650 $\rm \AA$ \\
 FEII\_UV\_EW\_ERR & DOUBLE & $\rm \AA$ & Uncertainties in REW\_FE\_2250\_2650  \\
 FEII\_OPT & DOUBLE[3] &   & Best-fit parameters for the optical \FeII\ model \\
 FEII\_OPT\_ERR & DOUBLE[3] &   & Uncertainties in the best-fit optical \FeII\ model \\
 FEII\_OPT\_EW & DOUBLE & $\rm \AA$ & Rest-frame equivalent width of optical \FeII\ within 4434-4684 $\rm \AA$ \\
 FEII\_OPT\_EW\_ERR & DOUBLE & $\rm \AA$ & Uncertainties in REW\_FE\_4434-4684 \\
 LOGL1350 & DOUBLE & [erg s$^{-1}$] & Continuum luminosity at rest-frame 1350 $\rm\AA$ \\
 LOGL1350\_ERR & DOUBLE & [erg s$^{-1}$] & Uncertainty in LOGL1350 \\
 LOGL1700 & DOUBLE & [erg s$^{-1}$] & Continuum luminosity at rest-frame 1700 $\rm\AA$ \\
 LOGL1700\_ERR & DOUBLE & [erg s$^{-1}$] & Uncertainty in LOGL1700 \\
 LOGL3000 & DOUBLE & [erg s$^{-1}$] & Continuum luminosity at rest-frame 3000 $\rm\AA$ \\
 LOGL3000\_ERR & DOUBLE & [erg s$^{-1}$] & Uncertainty in LOGL3000 \\
 LOGL5100 & DOUBLE & [erg s$^{-1}$] & Continuum luminosity at rest-frame 5100 $\rm\AA$ \\
 LOGL5100\_ERR & DOUBLE & [erg s$^{-1}$] & Uncertainty in LOGL5100 \\
\hline
   &   &  $\rm \AA$, $\rm \AA$, $10^{-17}$ erg s$^{-1}$ cm$^{-2}$, &  peak wavelength, 50\% flux centoid wavelength, flux,  \\
   &   & [erg s$^{-1}$], km s$^{-1}$, $\rm \AA$ &  log$L_{\rm line}$, FWHM, rest-frame equivalent width \\
 HALPHA & DOUBLE[6] & \nodata & For the entire \halpha\,profile (narrow and broad lines combined) \\
 HALPHA\_BR & DOUBLE[6] & \nodata & For the broad \halpha\,component \\
 NII6585 & DOUBLE[6] & \nodata & For the narrow [N\,{\sc ii}] \,$\lambda$6584 component \\
 HBETA & DOUBLE[6] & \nodata & For the entire \hbeta\,profile (narrow and broad lines combined) \\
 HBETA\_BR & DOUBLE[6] & \nodata & For the broad \hbeta\,component \\
 HEII4687 & DOUBLE[6] & \nodata & For the entire He\,{\sc ii}\,$\lambda$4687 profile (narrow and broad lines combined) \\
 HEII4687\_BR & DOUBLE[6] & \nodata & For the broad He\,{\sc ii}\,$\lambda$4687 component \\
 OIII5007 & DOUBLE[6] & \nodata & For the entire [O\,{\sc iii}] \,$\lambda$5007 profile \\
 OIII5007C & DOUBLE[6] & \nodata & For the core [O\,{\sc iii}] \,$\lambda$5007 profile \\
 CAII3934 & DOUBLE[6] & \nodata & For the Ca II K absorption line \\
 OII3728 & DOUBLE[6] & \nodata & \nodata \\
 NEV3426 & DOUBLE[6] & \nodata & For the entire Ne\,{\sc v}\,$\lambda$3426 profile (narrow and broad lines combined) \\
 MGII & DOUBLE[6] & \nodata & For the entire \MgII\,profile (narrow and broad lines combined) \\
 MGII\_BR & DOUBLE[6] & \nodata & For the broad \MgII\,component \\
 CIII\_ALL & DOUBLE[6] & \nodata & For the entire \CIII\,complex (\CIII, \SiIII, \AlIII) \\
 CIII\_BR & DOUBLE[6] & \nodata & For the broad \CIII\,component \\
 SIIII1892 & DOUBLE[6] & \nodata & \nodata \\
 ALIII1857 & DOUBLE[6] & \nodata & \nodata \\
 NIII1750 & DOUBLE[6] & \nodata & \nodata \\
 CIV & DOUBLE[6] & \nodata & \nodata \\
 HEII1640 & DOUBLE[6] & \nodata & For the entire He\,{\sc ii}\,$\lambda$1640 profile (narrow and broad lines combined) \\
 HEII1640\_BR & DOUBLE[6] & \nodata & For the broad He\,{\sc ii}\,$\lambda$1640 component \\
 SIIV\_OIV & DOUBLE[6] & \nodata & For the 1400 $\rm\AA$ complex \\
 OI1304 & DOUBLE[6] & \nodata & \nodata \\
 LYA & DOUBLE[6] & \nodata & \nodata \\
 NV1240 & DOUBLE[6] & \nodata & \nodata \\
 \hline
   &   & $\rm \AA$, $\rm \AA$, $10^{-17}$ erg s$^{-1}$ cm$^{-2}$, & Uncertainties in peak wavelength, 50\% flux centroid wavelength, flux,  \\
   &   & [erg s$^{-1}$], km s$^{-1}$, $\rm \AA$ & log$L_{\rm line}$, FWHM, rest-frame equivalent width \\
 HALPHA\_ERR & DOUBLE[6] & \nodata & \nodata \\
 HALPHA\_BR\_ERR & DOUBLE[6] & \nodata & \nodata \\
 NII6585\_ERR & DOUBLE[6] & \nodata & \nodata \\
 HBETA\_ERR & DOUBLE[6] & \nodata & \nodata \\
 HBETA\_BR\_ERR & DOUBLE[6] & \nodata & \nodata \\
 HEII4687\_ERR & DOUBLE[6] & \nodata & \nodata \\
 HEII4687\_BR\_ERR & DOUBLE[6] & \nodata & \nodata \\
 OIII5007\_ERR & DOUBLE[6] & \nodata & \nodata \\
 OIII5007C\_ERR & DOUBLE[6] & \nodata & \nodata \\
 CAII3934\_ERR & DOUBLE[6] & \nodata & \nodata \\
 OII3728\_ERR & DOUBLE[6] & \nodata & \nodata \\
 NEV3426\_ERR & DOUBLE[6] & \nodata & \nodata \\
 MGII\_ERR & DOUBLE[6] & \nodata & \nodata \\
 MGII\_BR\_ERR & DOUBLE[6] & \nodata & \nodata \\
 CIII\_ALL\_ERR & DOUBLE[6] & \nodata & \nodata \\
 CIII\_BR\_ERR & DOUBLE[6] & \nodata & \nodata \\
 SIIII1892\_ERR & DOUBLE[6] & \nodata & \nodata \\
 ALIII1857\_ERR & DOUBLE[6] & \nodata & \nodata \\
 NIII1750\_ERR & DOUBLE[6] & \nodata & \nodata \\
 CIV\_ERR & DOUBLE[6] & \nodata & \nodata \\
 HEII1640\_ERR & DOUBLE[6] & \nodata & \nodata \\
 HEII1640\_BR\_ERR & DOUBLE[6] & \nodata & \nodata \\
 SIIV\_OIV\_ERR & DOUBLE[6] & \nodata & \nodata \\
 OI1304\_ERR & DOUBLE[6] & \nodata & \nodata \\
 LYA\_ERR & DOUBLE[6] & \nodata & \nodata \\
 NV1240\_ERR & DOUBLE[6] & \nodata & \nodata \\
 \hline
 HA\_COMP\_STAT & DOUBLE[2] &   & Complex line pixel number, reduced $\chi^2$ \\
 HB\_COMP\_STAT & DOUBLE[2] &   & \nodata \\
 CAII3934\_LOC\_STAT & DOUBLE[2] &   & Local line pixel number, reduced $\chi^2$ \\
 OII3728\_LOC\_STAT & DOUBLE[2] &   & \nodata \\
 NEV3426\_LOC\_STAT & DOUBLE[2] &   & \nodata \\
 MGII\_COMP\_STAT & DOUBLE[2] &   & Complex line pixel number, reduced $\chi^2$ \\
 CIII\_COMP\_STAT & DOUBLE[2] &   & \nodata \\
 CIV\_COMP\_STAT & DOUBLE[2] &   & \nodata \\
 SIIV\_COMP\_STAT & DOUBLE[2] &   & \nodata \\
 LYA\_COMP\_STAT & DOUBLE[2] &   & \nodata \\
 LOGLBOL & DOUBLE & [erg s$^{-1}$] & Bolometric luminosity \\
 LOGLBOL\_ERR & DOUBLE & [erg s$^{-1}$] & Uncertainties in bolometric luminosity \\
 LOGMBH\_HB & DOUBLE & [M$_\odot$] & Single-epoch BH mass based on \hbeta \\
 LOGMBH\_HB\_ERR & DOUBLE & \nodata & Uncertainties in LOGMBH\_HB \\
 LOGMBH\_MGII & DOUBLE & \nodata & Single-epoch BH mass based on \MgII  \\
 LOGMBH\_MGII\_ERR & DOUBLE & \nodata & \nodata \\
 LOGMBH\_CIV & DOUBLE & \nodata & Single-epoch BH mass based on \CIV  \\
 LOGMBH\_CIV\_ERR & DOUBLE & \nodata & \nodata \\
 LOGMBH & DOUBLE & \nodata & Fiducial single-epoch BH mass \\
 LOGMBH\_ERR & DOUBLE & \nodata & \nodata \\
 LOGLEDD\_RATIO & DOUBLE &    & Eddington ratio based on fiducial single-epoch BH mass \\
 LOGLEDD\_RATIO\_ERR & DOUBLE &    & Uncertainties in LOGLEDD\_RATIO \\
 Z\_SYS\_LINE & DOUBLE[8] &   & Systematic redshift from individual lines \\
 & & & in the order of H$\beta_{\rm br}$, \OIII, \CaII\,3934, \OII\,3728, \\
 & & & \MgII, \CIII, \CIV, \SiIV \\
 Z\_SYS\_LINE\_ERR & DOUBLE[8] &   & Uncertainties in systematic redshift from individual lines 
\end{longtable*}



\section{Discussion}
\label{sec:disc}

\subsection{Comparisons with earlier measurements}
\label{sec:bh_mass}

We compare our spectral measurements with earlier studies on SDSS quasars \citep[e.g.,][]{Shen_etal_2011, Shen_etal_2019b,Rakshit_etal_2020} and generally find good agreement, despite minor differences in the exact fitting methods. One notable difference is that our global-fitting approach differs from the local-fitting approach in \citet{Shen_etal_2011}, and the measured \CIV\ FWHMs and fluxes are slightly different from the DR7Q measurements in \citet{Shen_etal_2011}.

\begin{figure*}
\centering
    \includegraphics[width=\linewidth]{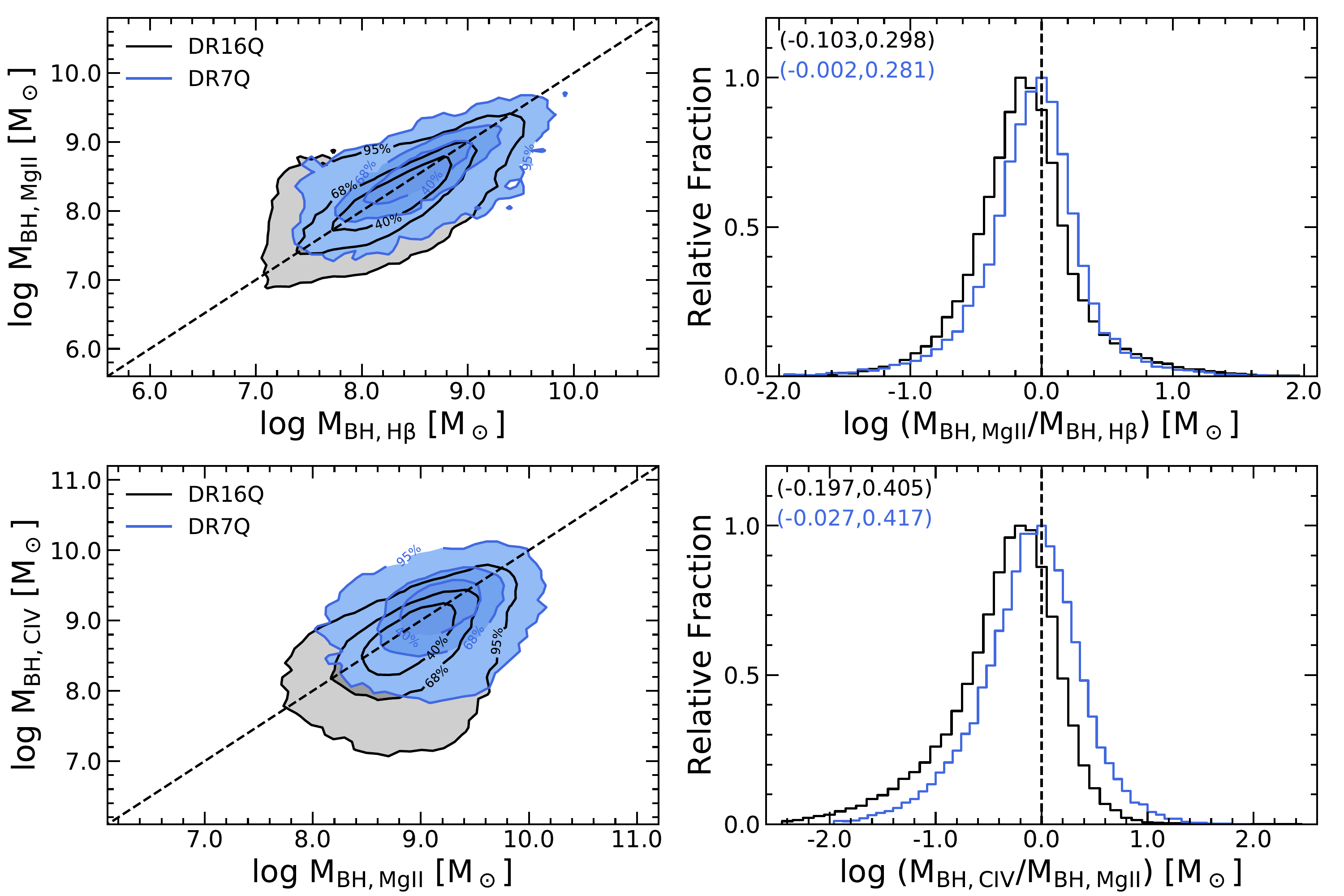}
   \caption{Upper panels: comparison between single-epoch virial black hole masses estimated using the broad \MgII\, line and broad \hbeta\ line. The mean ($\mu$) and dispersion ($\sigma$) of the mass difference distribution are noted on the top left of the right panel. 
    Lower panels: comparison between virial BH masses estimated from \CIV\ and \MgII. The contour levels correspond to the enclosed percentiles of the sample. While the BH masses are consistent between two line estimators for DR7 quasars \citep{Shen_etal_2011}, there are minor offsets in the mean BH mass between two estimators when extrapolating the same mass recipes to lower-luminosity DR16Q quasars. }
    \label{fig:MBH_comparison}
\end{figure*}

Another major difference is that the DR16Q sample probes much fainter luminosities than the DR7Q sample. While the fiducial BH mass recipes we adopted produce consistent results between two different lines for the luminous subset of DR7Q quasars, the BH mass consistency between two lines is degraded for lower-luminosity objects. In Fig.~\ref{fig:MBH_comparison} we compare virial BH masses between two lines covered in the same spectrum. For the DR7Q subset, we reproduce the results in \citet{Shen_etal_2011}. However, when extending to fainter immensities of DR16Q quasars, there is a mean offset of $\sim 0.1-0.2$ dex between two BH mass estimators. For many low-luminosity DR16Q quasars, we also find that the \CIV\ line has a rather narrow core, potentially arising from the narrow-line region or representing the intrinsic \CIV\ profile. The resulting \CIV\ FWHM is typically narrow and it is unclear if a narrow-line subtraction is necessary when estimating the \CIV-based BH masses.   

Because the systematic uncertainty for any of the single-epoch virial BH mass recipes is substantially larger ($\sim 0.4$\,dex) than typical measurement uncertainties, and because the extrapolation of these mass recipes calibrated with local reverberation mapping AGN samples to distant quasars is highly uncertain \citep[e.g.,][]{Shen_2013}, in this work we do not attempt to design new single-epoch mass recipes to bring all three estimators (\hbeta, \MgII, and \CIV) into agreement for DR16Q quasars. A more reasonable approach is to wait for better single-epoch BH mass recipes suitable for SDSS quasars, which are calibrated using direct reverberation mapping results of luminous quasars \citep[e.g.,][]{Shen_etal_2015a}.    




\subsection{Spectral diversity of quasars}

Broad-line quasars display a range of spectral diversities as well as correlations among physical properties. For example, the strength (equivalent width) of certain broad emission lines is correlated with quasar luminosity \citep[i.e., the Baldwin effect,][]{Baldwin_1977}, and many spectral properties correlate with the optical \FeII\ strength or Eddington ratio of the quasar, known as the Eigenvector 1 correlations \citep[e.g.,][]{Boroson_Green_1992,Sulentic_etal_2000,Marziani_etal_2001,Boroson_2002,Richards_etal_2011,Shen_Ho_2014}. The spectral measurements presented here for the DR16Q quasars can be used to perform detailed statistical studies of emission properties of quasars and correlations among them. We present a few examples below. 

\begin{figure}
\centering
    \includegraphics[width=\linewidth]{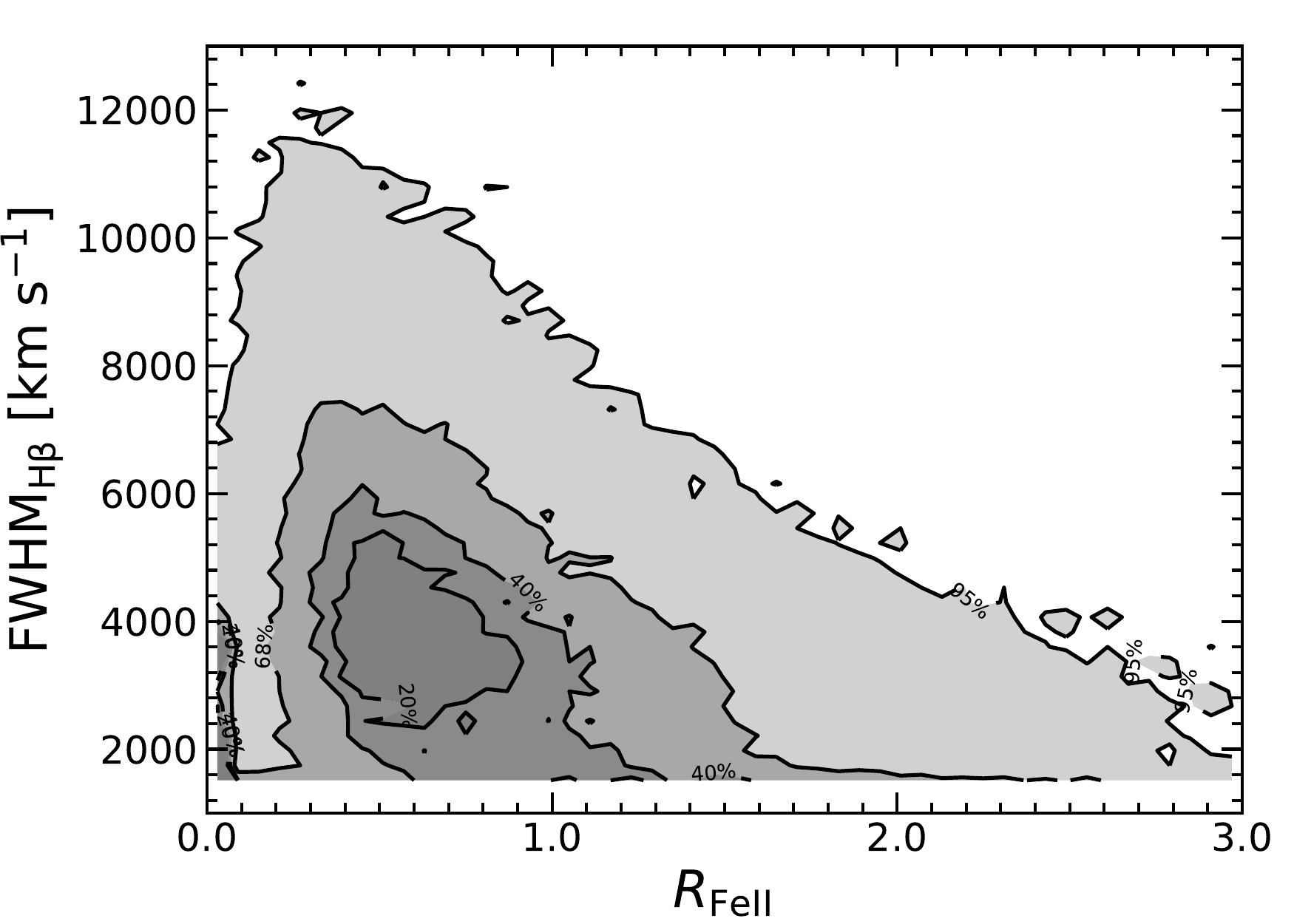}
    \caption{Distribution of DR16Q quasars in the plane of optical \FeII\ strength (denoted by $R_{\rm FeII}\equiv {\rm EW_{FeII\,4434-4684}}/{\rm EW_{H\beta,br}}$) versus broad \hbeta\ FWHM. }
    \label{fig:FWHM_HB_R}
\end{figure}

Fig.~\ref{fig:FWHM_HB_R} displays the distribution of quasars in the optical \FeII\ strength, defined by $R_{\rm FeII}\equiv {\rm EW_{FeII\,4434-4684}}/{\rm EW_{H\beta,br}}$, versus broad \hbeta\ FWHM plane. Such a plot was first seen in \citet{Boroson_Green_1992} and subsequently utilized in studies of the Eigenvector 1 relations of quasars \citep[e.g., the 4DE1 relations,][]{Sulentic_etal_2000}. A general trend of decreasing broad \hbeta\ width with increasing optical \FeII\ strength is seen, likely driven by increasing Eddington ratio as $R_{\rm FeII}$ increases \citep[e.g.,][]{Boroson_2002,Shen_Ho_2014,Sun_Shen_2015}. Orientation of the Balmer broad-line region may also play a role in the vertical dispersion of the distiribution \citep[e.g.,][]{Wills_Browne_1986,Marziani_etal_2001,Shen_Ho_2014}. 

\begin{figure}
\centering
    \includegraphics[width=\linewidth]{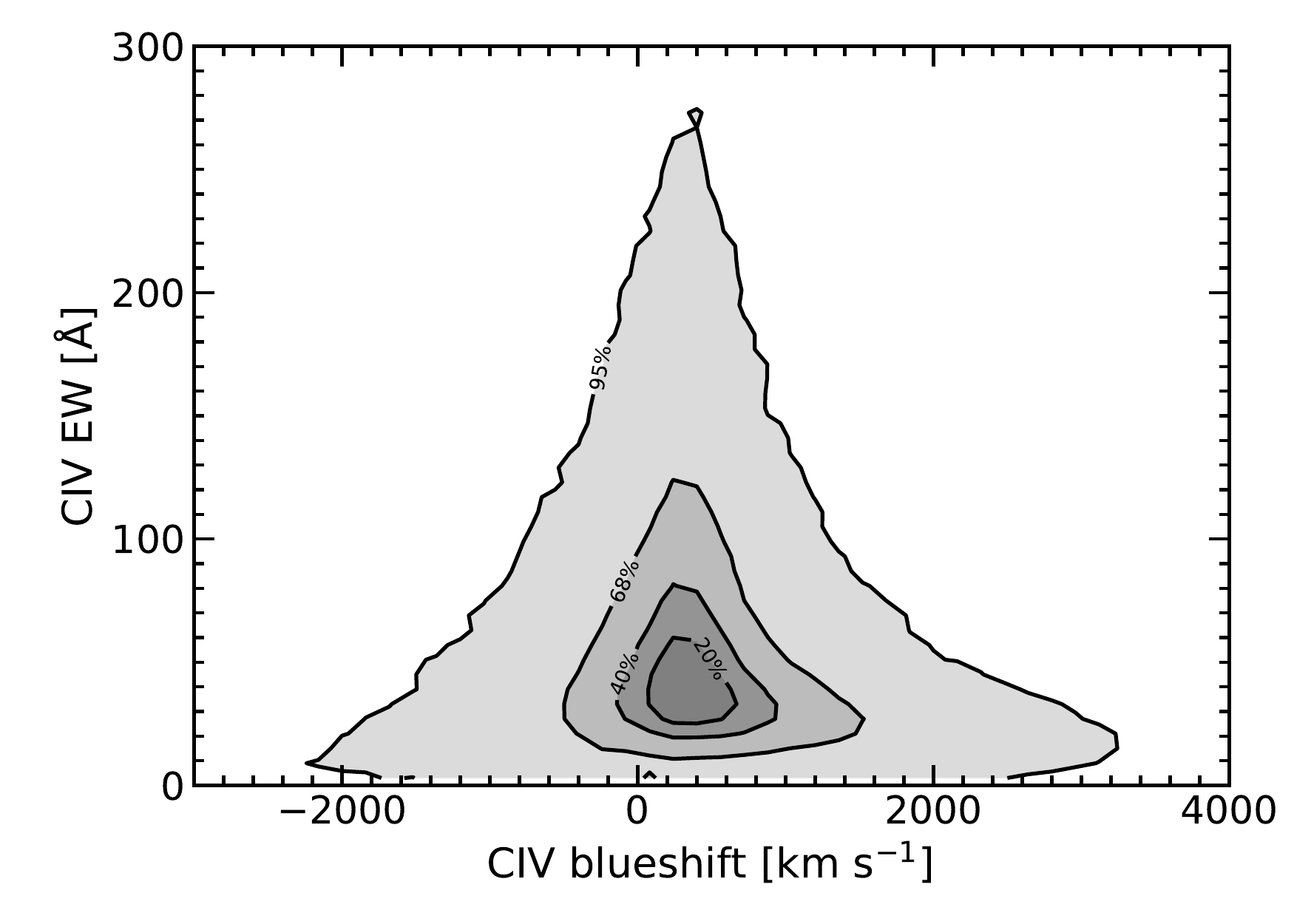}
    \caption{Distribution of DR16Q quasars in the \CIV\, equivalent width versus \CIV\ blueshift (wrt systemic) plane.}
    \label{fig:CIV_blueshift}
\end{figure}

In Fig.~\ref{fig:CIV_blueshift} we show the distribution of quasars in the \CIV\ blueshift versus equivalent width plane, where quasars occupy a similar wedge-shaped region \citep[e.g.,][]{Richards_etal_2011,Rivera_etal_2022}. The observed overall trend of decreasing \CIV\ EW with increasing \CIV\ blueshift is argued to be driven by accretion parameters \citep[e.g.,][]{Richards_etal_2011,Rivera_etal_2022}, which also correlates with multi-wavelength quasar properties, in particular X-ray properties \citep[e.g.,][]{Rivera_etal_2022}.



\section{Conclusions}\label{sec:con}

In this work we compile spectral measurements for 750,414 quasars included in the quasar catalog from SDSS DR16 \citep{Lyke_etal_2020}, using optical SDSS spectroscopy. Our final catalog includes continuum and line properties for a list of prominent lines in quasar spectrum, as well as derived quantities such as the black hole mass estimates based on single-epoch mass recipes and Eddington ratios.

This is a major update to the catalog for 105,783 SDSS-DR7 quasars presented in \citet{Shen_etal_2011}, with a more comprehensive line list and extension to lower quasar luminosities than the DR7 catalog. In particular, we present refined systemic redshifts and more realistic redshift uncertainties for these DR16Q quasars, demonstrating significant improvement over the original redshifts in DR16Q that were largely based on pipeline redshifts. 

All products from this work, including the catalog, input fitting parameter files, individual fits output, a python notebook demo, and update notes, can be accessed from \url{https://github.com/QiaoyaWu/sdss4_dr16q_tutorial}.  

\acknowledgments

This work is partially supported by NSF grant AST-2009947. QW acknowledges support from an Illinois Graduate Survey Science Fellowship. 

Funding for the Sloan Digital Sky Survey IV has been provided by the Alfred P. Sloan Foundation, the U.S. Department of Energy Office of Science, and the Participating Institutions. SDSS-IV acknowledges support and resources from the Center for High-Performance Computing at the University of Utah. The SDSS web site is www.sdss.org.


\bibliography{refs}

\end{document}